\documentclass[cup6b]{cupbook}
\usepackage{epsf}
\newcommand{\lSect}[1]{{\label{sec:#1}}}

\def\gtaprx {\lower .1ex\hbox{\rlap{\raise .6ex\hbox{\hskip .3ex
	{\ifmmode{\scriptscriptstyle >}\else
		{$\scriptscriptstyle >$}\fi}}}
	\kern -.4ex{\ifmmode{\scriptscriptstyle \sim}\else
		{$\scriptscriptstyle\sim$}\fi}}}
\def\ltaprx {\lower .1ex\hbox{\rlap{\raise .6ex\hbox{\hskip .3ex
	{\ifmmode{\scriptscriptstyle <}\else
		{$\scriptscriptstyle <$}\fi}}}
	\kern -.4ex{\ifmmode{\scriptscriptstyle \sim}\else
		{$\scriptscriptstyle\sim$}\fi}}}

\newcommand{\Msun}{{\ensuremath{\mathrm{M}_{\odot}}}}

\newcommand{\Sectff}[1]{{\ref{sec:#1}}}
\newcommand{\Sect}[1]{{Section~\Sectff{#1}}}

\begin{document}

\author[S.E. Woosley]{Stan E. Woosley\\ 
      Department of Astronomy and Astrophysics, \\
      University of California, Santa Cruz CA 95060}

\chapter{Models for Gamma-ray Burst Progenitors and Central Engines}

\section{Introduction}

For forty years theorists have struggled to understand gamma-ray
bursts (GRBs), not only where they are and the systematics of their
observed properties, but {\it what\/} they are and how they operate.
These broad questions of origin are often referred to as the problem
of the ``central engine''. So far, this prime mover remains hidden
from direct view, and will remain so until neutrino or
gravitational-wave signatures are detected. As discussed elsewhere in
this volume, there is compelling evidence that all GRBs require the
processing of some small amount of matter into a very exotic state,
probably not paralleled elsewhere in the modern Universe. This matter
is characterized by an enormous ratio of thermal or magnetic energy to
mass, and the large energy loading drives anisotropic, relativistic
outflows. The burst itself is made far away from this central source,
outside the star which would otherwise obscure it, by processes that
are still being debated (Chapters 7, 8).  The flow of energy is
modulated by passing through the star, which also explodes as a
supernova, and this modulation further obscures details of the central
engine.

The study of GRBs experienced spectacular growth after 1997 when the
first cosmological counterparts were localized (Chapter 4), and with
that growth in data came increased diversity. Still it is customary to
segregate GRBs into ``long-soft'' (LSBs) and ``short-hard'' (SHBs)
categories (Kouveliotou 1993)\nocite{Kou93}, though the distinction is
not always clear (Chapters 3 and 5; \Sect{hiz}).  Currently, it is
thought that most SHBs result from the merger of compact objects -
black holes and neutron stars - in galaxies and regions where the star
formation rate is low. There are exceptions, such as GRB\,050709, a
short hard burst that happened in a star forming galaxy (Covino
et~al.\ 2006)\nocite{Cov06}. But then Type Ia supernovae also happen
in spiral galaxies as well as ellipticals, and there is no reason to
expect the SHB progenitor population to be exclusively confined to old
galaxies (Prochaska et~al.\ 2006)\nocite{Pro06}, or even to galaxies
for that matter.  A few SHBs may be giant flares from soft gamma-ray
repeaters in galaxies that are relatively nearby (Palmer et~al.\ 2005;
Tanvir et~al.\ 2005)\nocite{Pal05,Tan05}. GRB\,070201 seems to have
happened in the Andromeda Galaxy (Mazets
et~al.\ 2008)\nocite{Mazet08}, but the event rate for compact mergers
there is $\sim$10$^{-5}$ - 10$^{-4}$ y$^{-1}$ (Kalogera
et~al.\ 2004)\nocite{Kal04}, and LIGO saw no signal (Abbott
et~al.\ 2006)\nocite{Abb08}.

Most LSBs, on the other hand, are clearly associated with massive star
formation and, since the lifetime of such stars is short, with massive
star death. This connection is strengthened by the observation that
many LSBs are accompanied by bright supernovae, when one might be
detectable (Chapter 9; Woosley \& Bloom 2006\nocite{Woo06a}). Most
LSBs {\it must} therefore be a consequence of neutron star or black
hole birth, and that means that LSBs are some variety of core-collapse
supernova. One of the greatest current challenges in the study of
stellar evolution is separating out the physical conditions that lead
to ordinary supernovae, which are much more frequent, and to LSBs.
Are they a continuum, or is there something uniquely different about
LSBs? It would help if we understood the mechanism of ``ordinary''
supernovae better. In fact, one of the greatest opportunities provided
by LSBs is the possibility of an improved understanding of massive
star death in general.

This chapter focuses on massive star models and is thus concerned
chiefly, though not exclusively, with LSBs. This does not imply that
massive stars are incapable of producing SHBs, only that the
connection with LSBs is more clearly demonstrated.  Indeed, the
observational distinction between LSBs and SHBs is becoming blurred.
For a recent review of SHBs see Nakar (2007)\nocite{Nak07}, and
\Sect{hiz}.

\section{Observational constraints}
\lSect{observations}

\subsection{The long-soft burst host environment and progenitor masses}
\lSect{hosts}

LSBs are not just extragalactic, they show a preference for high
redshift. Prior to {\it Swift\/}, the mean redshift of LSBs was 1.3;
now it is in the range of 2.2 (for 82 bursts, Jakobsson
et~al.\ 2006\nocite{Jak06}) to 2.6 (for 41 bursts with good redshift
determinations, Fiore et~al.\ 2007\nocite{Fio07}). Most bursts
observed with {\it Swift\/} originate from a time when the Universe
was only a few billion years old. This fixes LSBs to an epoch when
galaxies and metallicity were both evolving rapidly and the star
formation rate, at least in some galaxies, was high (Savaglio
et~al.\ 2008)\nocite{Sav08a}.  Some bursts do come from relatively
nearby galaxies, but most of these are subenergetic (Kaneko
et~al.\ 2007)\nocite{Kan07}. One possibility is that this reflects the
evolution of metallicity in the Universe (\Sect{Zobs} and
\Sect{Zmod}).

LSBs are concentrated in small, irregular galaxies and are strongly
correlated with the light in those galaxies, more so than Type II
supernovae (Fruchter et~al.\ 2006)\nocite{Fru06}. In fact, the
distribution of LSBs with light is similar to that of Type Ic
supernovae (Kelly, Kirshner and Pahre \ 2008)\nocite{Kel07} and, to
the extent that such supernovae are thought to originate from a very
massive population, LSBs probably come from stars more massive than 20
\Msun \ on the main sequence (Larsson et~al.\ 2007) \nocite{Lar07}.
Raskin et~al.\ (2008)\nocite{Ras08}, by constructing analytic models
of star-forming galaxies and the evolution of stellar populations
within them, find that the minimum progenitor mass on the main
sequence for LSBs is likely to be above 40\,\Msun.  \"Ostlin
et~al.\ (2008)\nocite{Ost08} find that the progenitor of GRB\,030329
was at least 25\,\Msun \ with only a small probability of models as
light as 12\,\Msun. Th\"one et~al.\ (2008)\nocite{Tho08} estimate that
the progenitor of GRB\,060505 had a mass of 32\,\Msun.

\subsection{Metallicity}
\lSect{Zobs}

Since line-driven and grain-driven mass loss is metallicity dependent
and mass loss removes angular momentum, one early prediction of the
models (\Sect{Zmod}) was that LSBs should be easier to produce when
the metallicity is low.  Low metallicity has been observed for several
local ($z<0.25$) LSB sites (Sollerman et~al.\ 2005; Stanek
et~al.\ 2006)\nocite{Sol05,Sta06} as well as in distant LSB hosts
(Fynbo et~al.\ 2003; Christensen et~al.\ 2004; Gorosabel et~al.\ 2005;
Fruchter et~al.\ 2006; Prochaska
et~al.\ 2008)\nocite{Fyn03,Chr04,Gor05,Fru06,Pro08}. Savaglio
et~al.\ (2009)\nocite{Sav08a}, in a study of LSB host galaxies at an
average redshift of $z=0.96$, found evidence for subsolar metallicity
in LSBs with an average metallicity of 1/6 solar for 17 of the hosts.
Chen et~al.\ (2008)\nocite{Che08} found evidence for a declining ISM
metallicity with decreasing galaxy luminosity in the star-forming
galaxy population at $z= 2 - 4$, i.e., the appropriate range for
LSBs. The average luminosity for 15 LSB hosts was about 10\% that of a
typical modern galaxy.

The broad-lined Type Ic supernovae that accompany LSBs have lower
metallicity than those that do not accompany LSBs (Modjaz
et~al.\ 2008)\nocite{Mod08}. This is particularly interesting, since
the other broad-lined SN Ic's at higher metallicity satisfy most of
the other observational constraints -- high mass progenitor, loss of
hydrogen envelope, and asymmetry -- required for GRBs.

Langer \& Norman (2006)\nocite{Lan06} estimate that 10\% of all stars
ever born were born with metallicity less than 10\% solar. The number
and redshift distribution of low metallicity stars is then a credible
match with what is seen for LSBs.  Campana
et~al.\ (2008)\nocite{Cam08} found evidence for (or at least
consistency with) a progenitor with rapid rotation and low metallicity
based on the X-ray spectrum of the surrounding material in
GRB\,060218. Specifically, the O/N and C/N ratios are difficult to
explain without rotational mixing and low metallicity.

All in all, the data are consistent with LSBs exhibiting a preference
for low metallicity regions, but still being possible, at least in a
mild form, when the metallicity is not very much less than solar (see
also Chapter 13).

\subsection{Energetics and beaming}
\lSect{energy}

Making an LSB requires not just a lot of energy, but directed
relativistic motion, i.e., a ``jet''. The existence of jets was
predicted by the models (Rhoads 1999)\nocite{Rho99}, which required an
asymmetric explosion to produce highly relativistic ejecta. This
prediction was later confirmed by observations of achromatic breaks in
the afterglow light curves (Fruchter et~al.\ 1999; Kulkarni
et~al.\ 1999; Harrison et~al.\ 1999; Stanek et~al.\ 1999; Frail
et~al.\ 2001)\nocite{Fru99,Kul99,Har99,Sta99,Fra01}.

An important constraint on the central engine is the total energy of
the burst and any accompanying supernova (\Sect{maxe}). This energy
includes both the non-relativistic ejecta, say $\beta \gamma \, \ltaprx
\, 2; \ v/c = \beta < 0.89$, and the relativistic ejecta. The
non-relativistic ejecta include the supernova and, by far, most of the
mass, but only the relativistic ejecta participate in making the LSB,
and do so with less than 100\% efficiency, perhaps much less. Energies
inferred for those few supernovae that have been clearly seen with
LSBs are typically a few times $10^{52}$ erg (e.g., Woosley \& Bloom
2006\nocite{Woo06a}), although aspherical explosions might work with
somewhat less energy (H\"oflich et~al.\ Wang 1999)\nocite{Hof99} for
supernovae that have not been studied far out on the tails of their
light curves.  This requirement on the energy comes from the need to
make a large mass of $^{56}$Ni, in order to make the supernova bright,
and, simultaneously, to provide a high velocity to a large mass of
ejecta.  Gamma-ray bursts with equivalent isotropic energies up to $9
\times 10^{54}$ erg have been reported (Abdo
et~al.\ 2009)\nocite{Abd09}, but the actual value is much less because
of beaming. A better constraint on the total relativistic energy comes
from analysis of the late time radio afterglow. Recent analysis of the
most energetic {\it Swift\/} bursts (Chandra et~al.\ 2008; Cenko
et~al.\ 2010a,b)\nocite{Cha08,Cen10a,Cen10b} suggests an upper bound
that is, again, in excess of $10^{52}$ erg. It is not known whether
these very energetic bursts also had comparably energetic supernovae,
but if they did, the total energy may, in some cases, be a substantial
fraction of 10$^{53}$ erg.

The relativistic component can also be quite weak. Li \& Chevalier
(1999)\nocite{Li99} found only $\sim10^{50}$ erg in the case of
GRB\,980425, and the matter may not have even been fully relativistic
in the sense of $\gamma \beta > $ 2 (Waxman \& Loeb
1999)\nocite{Wax99}. Given the large supernova energy in that event,
the energy in relativistic ejecta was less than 1\% that of the
non-relativistic ejecta, though still much greater than for ordinary
supernova. This contrasts with the $\sim$10$^{51}$ found in the
relativistic component of another supernova-related LSB, GRB 030329
(Kamble et~al.\ 2009)\nocite{Kam09}. There may also exist intermediate
phenomena between supernovae and LSBs with very little, but
non-trivial relativistic ejecta (Paragi et~al.\ 2010)\nocite{Par10}.

\section{Constraints from the models}

\subsection{Wolf-Rayet stars}
\lSect{WR}

Another early prediction of the models (Woosley 1993)\nocite{Woo93}
was that LSBs would only be produced by massive stars that had lost
their hydrogen envelopes, i.e., Wolf-Rayet stars of some sort. There
are two reasons the envelope must be lost. First, a relativistic jet
produced by any mechanism will not escape a hydrogenic star during the
characteristic duration of an LSB (Woosley
et~al.\ 2004)\nocite{Woo04}. The radius of even the most compact
progenitors of supernovae with hydrogen in their spectra (Type II) is
about 100 light seconds, and the head of the jet travels inside the
star substantially slower than light. It might be possible to have a
kind of very long transient if the central engine operated for several
minutes, but then it seems unlikely that a jet would break out after
such a long time and produce a burst that only only lasted a few
seconds. Probably such a jet, if one were ever made in a blue or red
supergiant, would die in the envelope producing an asymmetric
supernova and an X-ray transient (MacFadyen
et~al.\ 2001)\nocite{Mac01}.

A second problem with any sort of giant progenitor is that it implies
a slowly rotating helium core. If an envelope is present, the helium
core must have spun at a high rate for a very long time while embedded
in a medium that was essentially stationary. For current estimates of
magnetic torques (Spruit 2002)\nocite{Spr02}, the rotation of the iron
core when the star dies would be too slow to make an LSB (Heger
et~al.\ 2005)\nocite{Heg05}.  An obvious implication is that any
supernova seen in conjunction with an LSB must be Type I, not Type
II. This is consistent with observations so far.

\subsection{Rotation}
\lSect{rotation}

The existence of jets and the inability, so far, of isotropic models
powered by neutrino deposition to make very energetic supernova
explosions strongly suggest that the power source for most LSBs is
rotation.  That rotational energy, which ultimately is derived from
the gravitational contraction of the core, could take several forms -
the orbital motion of a disk or the rotational kinetic energy of a
neutron star or black hole - but it is the energy extracted from
rotation that powers the jet. While spherically symmetric shock
breakout (Colgate 1968; Chevalier \& Fransson 2008)\nocite{Col68,
  Chev08} can produce soft X-ray transients in WR-stars that might be
powered by neutrinos in a non-rotating model, the power and hardness
of an LSB spectrum requires much more energy per solid angle (Tan
et~al.\ 2001)\nocite{Tan01}.

That being the case, the preservation of angular momentum is
crucial. Since the LSB production rate is at least two orders of
magnitude less than the supernova rate, even of Type Ib/c (Bissaldi
et~al.\ 2007)\nocite{Bis07}, and since the required rotation rate
exceeds that of even the fastest solitary pulsars by an order of
magnitude, rotation is probably {\it the} governing factor that
determines whether a massive star death will produce an LSB. Efforts
to produce rapidly rotating systems fall into three categories:
interacting binaries, single star models with low magnetic torques,
and models based on the near homogeneous evolution of massive stars
with exceptionally high rotation rates on the main sequence.
 
Most binary merger and ``spin up'' models ignore magnetic torques in
the late stages of stellar evolution (e.g., Ivanova et~al.\ 2002; Joss
\& Becker 2007)\nocite{Iva02,Jos07}). They thus offer no obvious
advantage over single star models that also ignore magnetic torques
(Heger et~al.\ 2000; Hirschi et~al.\ 2005)\nocite{Heg00,Hir05} and
easily produce copious LSB progenitors. Indeed, two major problems
facing all models that ignore magnetic torques are the {\it
  overproduction} of LSBs and the lack of any clear alternative for
making the observed slow rotation rates of ordinary pulsars. Binary
models also usually overlook the large degree of {\it differential}
rotation that develops in LSB progenitors during their late stages of
evolution.  Studies by Petrovic et~al.\ (2005,
2006)\nocite{Pet05,Pet06} show that when magnetic torques are
included, mass transfer prior to helium ignition offers no clear
advantage over single star evolution.  Maintaining the surface
rotation rate of a Wolf-Rayet star (i.e., after the progenitor had
lost its hydrogen envelope) at a significant fraction of Keplerian
throughout helium burning would help mitigate the loss of angular
momentum by mass loss, and would therefore be beneficial. This might
be helpful in making some rotationally powered supernovae, or even low
energy LSBs at solar metallicity (\Sect{Zobs}), but even this would
require the accretion of helium or tidal locking in a very close
binary.

What is often overlooked in binary models for LSBs, especially those
that invoke tidal locking, is that the {\it deep core} must have
sufficient angular momentum to form a disk or a millisecond pulsar,
not just the surface of the star. The oxygen shell that bounds the
part of the core that collapses to a neutron star or black hole has a
typical radius of 2000 to 3000 \,km. If this shell rotated with a
period of 3 hours as in a rigidly rotating star in a very close,
tidally locked binary, the specific angular momentum at that shell
would be $j \sim 5 \times 10^{13}$ cm$^2$ s$^{-1}$, far too slow to
power either a collapsar or a millisecond magnetar.  If the disk did
not form from material inside the carbon-oxygen core, which has a
radius only about ten times bigger ($j \sim 5 \times 10^{15}$ cm$^2$
s$^{-1}$, but M $\gtaprx 10$\,\Msun), it is hard to see how the burst
could be accompanied by a supernova with the kind of spectrum and
light curve that have been observed in nearby events.  Still such
models (Tutukov \& Cherepashchuk 2003; Firmani et~al.\ 2004; Van den
Heuvel \& Yoon 2007)\nocite{Tut03,Fir04,Van07} are interesting for
giving sufficient angular momentum in the outer solar mass or so to
form a disk. Black hole formation in such systems would result in a
collapsar of a sort and may be common occurrences.  While this makes
them well worth exploring, the characteristic time scale for such
events would be much longer than for typical LSBs (\Sect{diversity}).

The second alternative is to admit the existence of magnetic torques,
but to question the accuracy of the Spruit (2002) formulation.
\nocite{Spr02}  Certainly it is difficult to know the strength of the
magnetic field generated by the dynamo action of instabilities in the
radiative layers of a massive star with any accuracy. It is agreed
that any purely poloidal field should suffer from instabilities on the
magnetic axis of the star (Tayler 1973)\nocite{Tay73}, but what is the
amplitude of these instabilities and can they drive a dynamo?  Even
more problematic is the dissipation rate one assigns to these fields,
once generated. Not surprisingly, there are diverse views on the
answer.  Braithwaite has carried 3D MHD calculations of the
Spruit-Tayler instability and finds results (Braithwaite 2006,
2008)\nocite{Bra06,Bra08} essentially consistent with
Spruit. Denissenkov \& Pinsoneault (2007)\nocite{Den07} on the other
hand find a greatly reduced magnetic viscosity, but their formulae
give a sun with a rapidly rotating core contrary to observations and
have been criticized (Spruit 2006)\nocite{Spr06}. Zahn, Brun and
Mathis (2007)\nocite{Zah07} see the instability, but do not see a
sustained dynamo action that could regenerate the mean fields.  Yoon
et al.\ (2008)\nocite{Yoo08} studied magnetic field creation in a
convective dynamo and found it not too much smaller in the radiative
regions outside the convective shells than the Spruit fields.  Suijs
et~al.\ (2008)\nocite{Sui08} took a more empirical approach, examining
the rotational velocity of white dwarfs generated in stellar evolution
calculations that include or do not include torques at the level
computed by Spruit. Only when the torques are included do they get
rotation rates slower than the observed upper bounds.  Ignoring
torques gives rotational rates two orders of magnitude bigger and
inconsistent with observed upper bounds, but there may be other ways
for white dwarfs to shed angular momentum (Charbonnel \& Talon
2005)\nocite{Cha05}. Clearly the last word has not been written here.

Assuming the approximate accuracy of the torques calculated by Spruit,
Heger et~al.\ (2005)\nocite{Heg05} find rotation rates for neutron
stars at birth that are within a factor of two of what is observed for
the youngest, most rapidly rotating pulsars (a still slower result
from the models would be preferable). They also find angular momenta
for the presupernova core of common supernovae that are too small to
power LSBs by either the collapsar model or the magnetar model
(\Sect{engine}), but perhaps large enough, in the more massive stars,
to make magnetars and rotationally powered supernovae. Again, a key
point is that the star must lose its hydrogen envelope before burning
much of its helium. However, one then becomes subject to the rapid
mass loss rates of WR-stars. An important recent development is the
realization that WR-star mass loss rates are very sensitive to
metallicity (\Sect{Zmod}). Another possibility is that WR mass loss
occurs preferentially along the rotational axis and therefore extracts
less angular momentum from the star (Meynet \& Maeder 2007; Georgy
et~al.\ 2008)\nocite{Mey07,Geo08}. It is probable that some
combination of removal of the envelope in a binary system, asymmetric
mass loss in the WR stage, and perhaps minor adjustment of the
parameters in the magnetic torque model can give stellar progenitors,
at death, with sufficient angular momentum to make an LSB by the
magnetar model, even at solar metallicity. A moderate decrease in
metallicity may even allow the existence of collapsars.

The third possibility is that the candidate stars are {\it very}
rapidly rotating from birth and that this dramatically changes their
evolution, even on the main sequence. The necessary rotational speeds
are over 300 km s$^{-1}$ and possibly 400 km s$^{-1}$ at the equator,
depending on metallicity. Such values are on the high-velocity tail of
what is observed. Stellar evolution studies show that such stars, due
to the operation of strong Eddington-Sweet circulation, have a novel
sort of ``quasi-chemically-homogeneous'' evolution (QCHE; Maeder
1987; Langer 1992; Yoon \& Langer 2005; Woosley \& Heger 2006; Yoon
et~al.\ 2006)\nocite{Mae87,Lan92,Yoo05,Woo06b,Yoo06}. They burn almost
all their hydrogen to helium on the main sequence and never expand to
become giant stars, going instead straight to the WR branch from the
main sequence. QCHE has been invoked to understand nitrogen-rich and
helium-rich massive main sequence stars in the Magellanic Clouds (see
refs in Yoon et~al.\ 2006\nocite{Yoo06}).  A combination of models is
also possible in which accretion from a binary companion spins up a
star to the point where it experiences QCHE (Cantiello
et~al.\ 2007)\nocite{Can07}.

Because the large density contrast between helium core and envelope
never develops in stars that experience QCHE, the core is not spun
down nearly so much during helium burning.  Even with the inclusion of
standard Spruit magnetic torques, the iron cores of such stars rotate
very rapidly. The two groups who work on this sort of model (Woosley
\& Heger 2006; Yoon et~al.\ 2006)\nocite{Woo06b,Yoo06} find that a
moderate decrease in metallicity (to reduce WR mass loss) results in
LSB progenitors with sufficient angular momentum to make a
collapsar. Yoon et~al.\ concluded that LSB production would occur for
metallicities Z $<$ 0.004 so that 50\% of all LSBs would happen beyond
redshift z = 4. That estimate is probably overly restrictive. LSBs
might be made by magnetars, which require less angular momentum than
collapsars, and the distribution of iron in low mass galaxies with
redshift may be different than that of oxygen in high mass
galaxies. The observability of distant bursts in a flux-limited sample
could also complicate matters.

A final possibility that could account for a small fraction of LSBs is
that the progenitor does not directly involve the death of a massive
star.  An example could be black holes merging with white dwarfs
(Fryer et~al.\ 1999)\nocite{Fry99b}. There the amount of angular
momentum is so large that an unusual kind of burst is predicted,
probably a very long one.

\subsection{Metallicity}
\lSect{Zmod}

Even on the main sequence, stars of lower metallicity are found to
rotate more rapidly (Maeder \& Meynet 2001; Ekstr\"om
et~al.\ 2008)\nocite{Mae01,Eks08}. Observations suggest that
differences in rotation rate persist all the way to the zero age main
sequence (Martayan et~al.\ 2007),\nocite{Mar07} suggesting that low
metallicity stars may also be {\it born} rotating faster, but see also
(Peng et~al.\ 2005)\nocite{Pen04}.  Reducing the mass loss also
results in a star that, at death, is heavier and thus more prone to
making a black hole, a requisite for the collapsar model. Extremely
low metallicity may also alter the initial mass function, producing
heavier stars that would frequently make black holes (e.g., Yoshida
et~al.\ 2006\nocite{Yos06}).

It is important to note the key role of the element iron here. Mass
loss in red giant stars is dependent on dust grain formation, which
will also be suppressed at low metallicity, but mass loss during the
red giant stage is not so relevant here. Unless the complete envelope
is removed quickly, any reasonable residual envelope will brake the
rotation of the core. It is the large number of iron lines that provide
the opacity for WR-star winds. The initial carbon and oxygen
abundances, and even the larger abundances of these elements made by
the star itself do not matter, unless the iron abundance is so low
that mass loss would be negligible (Vink \& de Koter
2005)\nocite{Vin05}.  Frequently, astronomers measure other elements
from which they derive ``metallicity'', but the nucleosynthetic
history of iron is different from that of oxygen and at low
metallicity, the iron to oxygen ratio might be as much as four times
less than solar (Nissen et~al.\ 2002)\nocite{Nis02}.  Of course, it
should go without saying that the iron abundance that matters is the
one for the star that blows up, not some other part of the galaxy that
it died in.

One does not expect the metallicity cutoff to be a sharp one, but as
the metallicity increases, the collapsing cores of massive stars will
tend to have less mass and angular momentum. All else being equal,
this would imply that the bursts seen closer by, presumably with
higher average metallicity, would be less energetic, as seems to be
the case. Of course, it is easier to discover weaker bursts that are
nearby, but analysis of {\it Swift\/} data by Kocevski \& Butler
(2008)\nocite{Koc08} suggest that nearby bursts are less energetic.

In the collapsar model, one would also expect that decreasing
metallicity would increase the reservoir of material available to form
a disk and thus increase both the duration and total energy of the
jet.  So far searches for such a correlation have been inconclusive,
and suggest that the duration may actually {\it decrease} with red
shift (Wei \& Gao 2003\nocite{Wei03}; though see \Sect{hiz}).

\subsection{Mass}
\lSect{mass}

If LSBs accompany neutron star or black hole birth, they must come
from stars with main sequence masses of at least 8\,\Msun. Since the
helium and heavy element shells of presupernova stars in the
8--12\,\Msun\ range have little mass, it would be difficult to produce
more than a trace of $^{56}$Ni in any explosion, so any supernova
would be faint. Such light stars also do not lose their envelopes
unless they are in a close binary and they have long evolutionary
times during which magnetic torques could slow their rotation. It
seems a safe bet that LSBs come from stars considerably more massive
than 10\,\Msun, a fact consistent with observational limits
(\Sect{hosts}).

Much beyond that though, theory offers uncertain guidance. In the QCHE
models (\Sect{rotation}), rotationally-induced mixing leads to
progenitors almost entirely composed of oxygen that are not much
smaller than the star's mass on the main sequence, especially if the
metallicity is low. A 16\,\Msun \ main sequence star with very rapid
rotation might lead to a 14\,\Msun \ presupernova model capable of
making a collapsar (Woosley \& Heger 2006)\nocite{Woo06b}. Without
QCHE, a 14\,\Msun \ helium core would have required a 35\,\Msun \ main
sequence star (Woosley \& Weaver 1995)\nocite{Woo95}, so the mixing
has a big effect.

The gravitational binding energy of a presupernova star increases
rapidly with the mass of the helium core and for masses over $\sim8$
\Msun \ (i.e., the helium core of a 25\,\Msun \ star evolved without
rotation) exceeds $10^{51}$\,erg (Woosley
et~al.\ 2002)\nocite{Woo02}. This makes it harder for the star to
explode, and makes it more likely that some matter will fall back in
those that do explode. Shallower density gradients near the iron core
also result in a larger accretion rate on the proto-neutron star
during the fist few seconds of the explosion, increasing the
likelihood of a ``failed'' explosion in which a black hole
forms. Fryer (1999)\nocite{Fry99a} estimated that black holes would
form during the deaths of non-rotating main sequence stars heavier
than 20\,\Msun \ from fall back, and would form promptly for stars
above 40\,\Msun. Several caveats are in order though. First, as
mentioned, rotation substantially alters the relation between helium
core mass and main sequence mass. Fryer's 40 \Msun \ limit could
correspond to a rapidly rotating main sequence star of only
20\,\Msun. Second, his calculations assumed the physics and
(two-dimensional) codes of 1995. Many calculations since have given
diverse results, and the definitive studies in three dimensions have
yet to be done, even without rotation. Finally, it may not be
necessary to make a black hole to generate a LSB. The magnetar model
obviously does not require this. However, in the magnetar model, LSBs
with bright supernova still require a massive progenitor to make
adequate $^{56}$Ni (Nomoto et~al.\ 2005, 2007)\nocite{Nom05,Nom07}.

Theory also suggests that too {\it big} a mass could actually impede
the production of an LSB (Yoon et~al.\ 2006; Woosley \& Heger
2006)\nocite{Yoo06,Woo06b}. This is because more massive WR stars are
currently thought to lose mass at a greatly accelerated rate that more
than compensates for their reduced lifetime. The maximum value depends
upon what is assumed about metallicity and uncertain mass loss rates,
but the limit is larger at lower metallicity.  In very massive stars
that produce presupernova helium cores over 45\,\Msun, the pulsational
pair instability is encountered (\Sect{ppsn}). For QCHE and no mass
loss, the limit could be as low as 45\,\Msun \ on the main sequence.

Taken together, the observational limits on mass are consistent with
those from theory, but it would not be surprising to find an LSB
coming from a star with a main sequence mass of 15\,\Msun, or even
less if the rotation rate were high and the metallicity (i.e., WR mass
loss rate) very low (Yoon et~al.\ 2006)\nocite{Yoo06}.

\section{Central engines -- A tale of two models}
\lSect{engine}

The association of LSBs with massive star death has rekindled an
age-old debate regarding the mechanism by which supernovae explode.
Since the 1930's it has been realized that supernovae are a
consequence of gravitational collapse in the cores of highly evolved
massive stars (Baade \& Zwicky 1934; Oppenheimer \& Volkov
1939)\nocite{Baa34,Opp39}, but there are three ways this might come
about. The iron core may collapse to a neutron star that is slowly
rotating and make an explosion powered by neutrinos (Colgate \& White
1966)\nocite{Col66}, or it may make a rapidly rotating neutron star
whose rotation powers the explosion by magnetic processes (Ostriker \&
Gunn 1971)\nocite{Ost71}. More recently it has been realized that
black hole formation can also power energetic explosions if the star
has sufficient rotation to form a disk around the hole (Woosley
1993)\nocite{Woo93}.

The necessary angular momentum in the rotational models is much larger
than is observed in common pulsars, suggesting that either pulsars are
somehow braked during (though see Ott et~al.\ 2006\nocite{Ott06}) or
after (though see Heger et~al.\ 2005\nocite{Heg05}) their birth, or
that only a small fraction of supernovae are powered this way. Recent
multi-dimensional simulations that ignore rotation have been
successful in blowing up at least the lightest supernovae with just
neutrinos (Janka et~al.\ 2007; Burrows et~al.\ 2007a; Messer
et~al.\ 2008)\nocite{Jan07,Bur07,Mes08}, and it is a reasonable
expectation that 15 to 20\,\Msun \ progenitors might also explode if
the calculation is done carefully in three dimensions. This would be
enough to explain the majority of ordinary supernovae and would be
consistent with pre-supernova model calculations (\Sect{rotation})
that give slowly rotating cores for pre-supernova stars in this mass
range. Since LSBs are rare, they are not constrained by pulsar
statistics and may come from unusual stellar evolution. Moreover, the
rapidly rotating compact object may either be a black hole or have
such a strong magnetic field that any remaining pulsar is rapidly
braked.

\subsection{The millisecond magnetar model}

A neutron star born with a rotation rate of $\sim$1\,ms (the maximum
rate without large scale deformation and rapid braking by
gravitational radiation) contains a large amount of energy, $E = 0.5 I
\Omega^2 / \sim 3 \times 10^{52}$ erg for a moment of inertia I = 80
km$^2$\,\Msun \ (Lattimer \& Prakash 2007)\nocite{Lat07}. The dipole
spin down luminosity, $L \sim B^2 R^6 \Omega^4/c^3 \sim 10^{50}$ erg
s$^{-1}$ for $B \sim 10^{15}$ Gauss, is typical of LSBs and the
supernovae that accompany them. Magnetars (Thompson \& Duncan
2006)\nocite{Tho96} are known to have a magnetic field strength close
to what is needed (Woods 2008)\nocite{Wood08}. They are found in young
supernova remnants and are observed to be spinning down rapidly. They
may be surrounded by disks from fallback debris in the explosion (Wang
et~al.\ 2006)\nocite{Wan06}, but are not powered by these disks. Their
birth rate in our Galaxy is estimated to be 0.1 per century
(Kouveliotou et~al.\ 1998)\nocite{Kou98} or more (Gill \& Heyl 2007;
Muno et~al.\ 2008)\nocite{Gil07,Mun08a}, and they seem to be derived
from a stellar population that may be heavier than 30\,\Msun\ (Muno
et~al.\ 2006; Muno 2008)\nocite{Mun06,Mun08b}, though see Davies
et~al.\ (2009)\nocite{Dav09}. Given that their birth rate is not too
much smaller than that of ordinary supernovae, only a small fraction
($\sim1$\%) of all magnetars can make bright LSBs when they are
born. The birth of the rest could contribute to supernova energetics
and asymmetry much more frequently, however, and only a subset with
particularly high magnetic field and rotation rate in stars without
hydrogen envelopes might make LSBs. It is interesting in this regard
that stellar evolution calculations predict that neutron stars with
more rapid rotation rates are made in more massive stars (Heger
et~al.\ 2005)\nocite{Heg05} and that the distribution of LSBs in their
host galaxies suggests a correlation with very massive stars (Wachter
et~al.\ 2008\nocite{Wac08}; \Sect{hosts}).

Recent models of millisecond magnetars show that they are capable of
producing relativistic outflows (Bucciantini et~al.\ 2008; Komissarov
\& Barkov 2008)\nocite{Buc08,Kom08} and might be the source of LSBs
(Usov 1992; Duncan \& Thompson 1992; Thompson
et~al.\ 2004)\nocite{Uso92,Dun92,Tho04}. A compelling case can be made
that magnetar activity may be involved in producing a significant
fraction of supernovae as well. So long as rapid rotation (P $<$ 5
ms) is required to create magnetar-like magnetic fields (Duncan \&
Thompson 1992, 1996)\nocite{Dun92,Dun96}, a rotational energy $\gtaprx
10^{51}$ erg must be dissipated by the neutron star. If that energy
does not come out as neutrinos or gravitational radiation, then it
must be an appreciable contribution to the supernova's kinetic
energy. There are two ways out of this. The iron core may not rotate
this fast when it collapses and contracts to $R \sim 10$\,km. This is
apparently the case for the roughly 90\% of supernovae that do {\it
  not} make magnetars, but make ordinary pulsars
instead. Alternatively, the energy of the rotation might be thermally
dissipated into neutrinos. Calculations so far (Dessart
et~al.\ 2008)\nocite{Des08} do not show this happening.

Despite these favorable arguments, it has not yet been demonstrated
that a magnetar model can produce an LSB. In some calculations, the
rotational energy is dissipated in sub-relativistic outflow (i.e., the
supernova) leaving little to power the burst (Dessart
et~al.\ 2008)\nocite{Des08}. In others a prior supernova (Bucciantini
et~al.\ 2008)\nocite{Buc08} or initial conditions that readily make a
supernova (Komissarov \& Barkov 2008)\nocite{Kom08} are assumed in
order to relieve the accretion onto the pulsar. This is not
surprising. A fully MHD 3D simulation of core collapse including
realistic neutrino transport and high density physics will be
hard. But it will be necessary because neutrinos alone seem unlikely
to launch a successful shock in the same very massive stars where they
have failed for so long.  The model must also satisfy some rather
tight constraints on energetics and nucleosynthesis (\Sect{ni56}).

If magnetars make LSBs, then their jets are likely to be Poynting flux
dominated. Neutrino energy deposition would not make a jet as in early
versions of the collapsar model (MacFadyen \& Woosley
1999)\nocite{Mac99}, and the current models suggest magnetically
dominated jets (Bucciantini et~al.\ 2008, 2009)\nocite{Buc08,Buc09}.

\subsection{The collapsar model}
\lSect{collapsar}

Supernovae must occasionally, and perhaps frequently, leave black hole
remnants. We see black holes, or their effects, in binary systems and
apparently some of them rotate very rapidly (McClintock et~al.\ 2006;
2007; Liu et~al.\ 2008)\nocite{Mac06,Mac07,Liu08}. In some cases,
these black holes are quite massive (Orosz
et~al.\ 2007)\nocite{Oro07}, suggesting the collapse of the entire
helium core. It is believed that the rapid spin and large masses
reflect natal properties, not the effects of accretion in a binary.

A black hole can be formed in a supernova either promptly or by fall
back. If it is by fall back, then the loosely bound hydrogen envelope
may be ejected in the process. If not, in the absence of rotation, the
entire star must collapse. For decades, calculations of iron-core
collapse in non-rotating massive stars over about 25\,\Msun \ formed
black holes promptly (Fryer 1999; Buras
et~al.\ 2006)\nocite{Fry99a,Bur06}, though the defining 3D studies
have yet to be done. It is thus a reasonable, if unproven, assumption
that some massive stars form black holes without initially developing
violent explosions in their cores. That being the case, a black hole
and an accretion disk must ultimately come to exist, since the outer
layers, even of red supergiants, have too much angular momentum to
fall into the hole directly. Several studies now show that black holes
experiencing very rapid, optically thick accretion will produce
relativistic outflows (Komissarov 2001; McKinney \& Gammie 2004;
Hawley \& Krolik 2006, Barkov \& Komissarov 2008; Tchekhovskoy
et~al.\ 2008)\nocite{Kos01,Mck04,Haw06,Bar08a,Tch08}. It is thus
reasonable to expect that some kind of luminous transient will
accompany the prompt birth of a black hole. The transients need not
all be ordinary gamma-ray bursts however. The free fall time of red
supergiant envelopes is hours to days. The observable signal might
more closely resemble an ultra-luminous X-ray source (Li
2003)\nocite{Li03} or even a blazar (\Sect{diversity}).

If the inner core rotates more rapidly though, and a black hole is
formed promptly, an LSB can result (Woosley 1993; Woosley \& MacFadyen
1999)\nocite{Woo93,Mac99}.  The angular momentum needed is at least
the value of the last stable orbit around a black hole of several
solar masses, $j = 2 \sqrt{3} GM/c = 4.6 \times 10^{16} (M_{\rm
  BH}/3\Msun)$ cm$^2$ s$^{-1}$ for a non-rotating hole and $j =
2/\sqrt{3} GM/c = 1.5 \times 10^{16} (M_{\rm BH}/3\Msun)$ cm$^2$
s$^{-1}$ for a spinning black hole with Kerr parameter $a=1$.  This
compares with an angular momentum needed for the millisecond magnetar
model of $j = R^2 \Omega \sim 5 \times 10^{15}$ cm$^2$ s$^{-1}$ (if
$\Omega \sim 5000$ rad s$^{-1}$ and R = 10 km). Since the black holes
in the collapsar model are typically very rapidly rotating and since
the specific angular momentum at 3 \Msun \ is greater than that at
1.5\,\Msun, the minimum angular momentum requirements of the collapsar
and millisecond magnetar models are comparable, though making a
millisecond magnetar is clearly easier. Models exist that would give
the necessary angular momentum for either (\Sect{rotation}) in a star
that has lost its hydrogen envelope before dying.

Two other necessary conditions for the collapsar model are that the
star not explode prematurely truncating black hole formation, and that
the jet, once produced, escapes the star. Considerable progress has
been made exploring the second requirement. Calculations (Aloy
et~al.\ 2000; Zhang et~al.\ 2004)\nocite{Alo00,Zha04} have shown that
an energy-loaded relativistic jet injected near the center of a
massive star will penetrate the star and produce a streaming jet with
the necessary Lorentz factor and opening angle to make an LSB. The
distribution of jet energy with angle may depend upon the mass of the
progenitor star with more massive progenitors having broader jets
(Mizuta \& Aloy 2009)\nocite{Miz09}. In calculations so far, the
energy loading was assumed to be thermal and the calculation ignored
magnetic fields, but more recent calculations of Poynting flux jets
show that they too can achieve a large terminal Lorentz factor
(Tchekhovskoy et~al.\ 2008)\nocite{Tch08}.  This is important since
many calculations (Komissarov 2001)\nocite{Kos01} now show that the
Blandford-Znajek (1977)\nocite{Bla77} mechanism, and other MHD
processes (Komissarov 2001; McKinney et~al.\ 2004; Hawley \& Krolik
2006)\nocite{Kos01,Mck04,Haw06} are probably the origin of
relativistic jets. These mechanisms work well in the context of the
collapsar model (Barkov \& Komissarov 2008; Tchekhovskoy
et~al.\ 2008)\nocite{Bar08a,Tch08}, but the jets are Poynting
dominated.  Neutrino annihilation as originally discussed in
(MacFadyen \& Woosley 1999; Popham et~al.\ 1999)\nocite{Mac99,Pop99}
is probably not adequate (Nagataki et~al.\ 2007)\nocite{Nag07} or at
least not as important as MHD processes (McKinney 2005a)\nocite{Mck05}
in driving jets. The relative contributions of neutrinos and MHD
processes may depend on the black hole mass, spin rate, and accretion
rate though. For black holes larger than a few \Msun, neutrino
annihilation is probably not a major effect. For lower masses, hybrid
models may exist in which neutrino annihilation helps boost or
initiate an MHD jet. It may also be that the jet is more thermal in
nature while it stays inside the star, but more MHD like after the
polar axis has been evacuated.

In addition to the jet itself, the ``disk wind'' (MacFadyen \& Woosley
1999; Kohri et~al.\ 2005; McKinney \& Nareayan 2007; Barkov \&
Komissarov 2008)\nocite{Mac99,Koh05,Mck07,Bar08a} plays an important
role in the collapsar model and shows promise of providing the
$\sim10^{51}$ erg of energy input to a large solid angle needed to
blow up the star and to make the necessary $^{56}$Ni (MacFadyen \&
Woosley 1999; Nagataki et~al.\ 2007)\nocite{Mac99,Nag07}. It is
important to note here that the properties of the supernova that
accompanies an LSB are given by the disk wind, not the jet itself. The
explosion energy and brightness of the supernova are, to some extent,
decoupled from the properties of the burst. By disk wind here, we also
mean to include the funnel of subrelativistic mass ejection
surrounding the jet in the black-hole spin powered model.

There is thus no compelling reason why the LSB jet energy, the
supernova kinetic energy, or the $^{56}$Ni yield should be a constant.
Faster rotating black holes may accelerate relativistic jets more
efficiently. Hawley \& Krolik (2006)\nocite{Haw06} estimate the
efficiency for producing the jet scales as $0.002/(1 \ - \ a)$ with
$a$ the Kerr parameter (see also McKinney 2005b\nocite{Mck05b}). For
accretion rates of 0.01\,\Msun \ s$^{-1}$ and $a$ = 0.9, this is about
$4 \times 10^{50}$ erg s$^{-1}$. Given that lower metallicity
progenitors end up with more rapidly rotating cores, this efficiency
factor suggests stronger bursts at lower metallicity. Hawley \& Krolik
also calculate that a substantial fraction of the accretion energy
ends up in a funnel-like sub-relativistic outflow surrounding the
jet. This plus the disk wind may be important to the dynamics of the
supernova.

One also expects the power of the burst to depend on the duration and
rate of accretion. For a simple dipole case, the efficiency of
extracting black hole rotational energy may go as $\dot M^{1/2}$
(Haley \& Krolik 2006)\nocite{Haw06}, and of course accreting more
total mass will make more energy. This again favors more massive stars
with higher rotation rates (so that more of the star forms a disk). It
has been suggested that collapsars with low accretion rate make less
$^{56}$Ni (Lopez-Camara et~al.\ 2008)\nocite{Lop08}.

A severe concern for the collapsar model though is the poorly explored
formation process for the black hole in a situation where enough
rotation exists to form a disk.  As with magnetar birth, this is a
hard problem involving neutrino physics, MHD, high density physics,
and general relativity all coupled to three-dimensional hydrodynamics.
A preliminary attempt has been made by Dessart
et~al.\ (2008)\nocite{Des08} who conclude, for conditions where the
collapsar is expected to work, that a magneto-rotational instability
(MRI) may blow up the star before a black hole is formed. Their
results are sensitive to an approximate treatment of the magnetic
field evolution though (Etienne et~al.\ 2006)\nocite{Eti06}. Using the
field of the actual pre-supernova model, they do make a black
hole. Using a field that they postulate would be created by a well
resolved MRI, they get a powerful explosion. It is also troubling that
the MRI that plays such a dominant role in the collapse is completely
ignored in the stellar evolution just before the collapse. With such
powerful radial fields, $B_r \sim B_{\phi}$, the core would rotate
more slowly. In fact, since their model uses most of its energy making
a supernova little is left to make an LSB jet by {\it any}
mechanism. Still this is a troubling gap in an otherwise successful
model.

\section{A diverse set of phenomena}
\lSect{diversity}

Given that the death of a massive star, with variable mass, angular
momentum, and structure, somehow, sometimes creates a powerful jet
near its center, a great variety of phenomena are possible. The LSBs
and X-ray flashes studied so far may just be the most easily
recognized consequences.

\subsection{Core-collapse supernovae} 

Polarization data suggest the breaking of spherical symmetry in {\it
  all} types of core-collapse supernovae (Wang \& Wheeler
2008)\nocite{Wan08}. For Type II supernovae, the deformation becomes
greater as one peers, with time, deeper into the explosion. This does
not necessarily imply that the typical supernova explosion is powered
by rotation, but it does imply strong symmetry breaking either during
or shortly after the explosion. This symmetry breaking could be due to
the rotation of the neutron star, vibrational energy input by
asymmetric fallback, or accretion instabilities during fallback
(Scheck et~al.\ 2006; Blondin \& Mezzacappa 2007; Burrows
et~al.\ 2007a)\nocite{Sch06,Blo07,Bur07}. In the future, it will be
important to understand the full continuum of massive star death. Are
LSBs a separate phenomenon or, as seems more likely, just the
extremity of a distribution of asymmetric supernova explosions?  What
are the necessary mass and rotation rate of stars for which rotation
{\it is} the dominant source of explosion energy?  Numerical
simulation may resolve this issue in the next decade.

\subsection{Magnetar birth} 

Magnetars with field strengths 10$^{14}$ to 10$^{15}$ Gauss exist and
may be formed in the birth of 10\% of all neutron stars (Kouveliotou
1998)\nocite{Kou98}. This is far too many for every magnetar birth to
produce an LSB. What do the rest look like? Recently, it has been
suggested (Woosley 2009; Kasen \& Bildsten 2010)\nocite{Woo09,Kas09}
that the emission of young magnetars might power the light curves of
many kinds of supernovae, including ultra-luminous ones and perhaps
some of the supernovae associated with LSBs. Interestingly, the
magnetars with the strongest fields spin down the quickest and
contribute less to the later light curve. Finding just one supernova
whose light curve is unambiguously due to magnetar energy deposition
would strengthen the case for magnetar involvement in
LSBs. Unfortunately, unique diagnostics may be hard to find. The very
late time light curve might distinguish between a $^{56}$Ni power
source and a pulsar in a Type Ibc supernova. There may also be
spectroscopic differences depending on how the magnetar deposits its
energy.

\begin{figure}
\begin{center}
 \epsfxsize=10cm \epsfbox{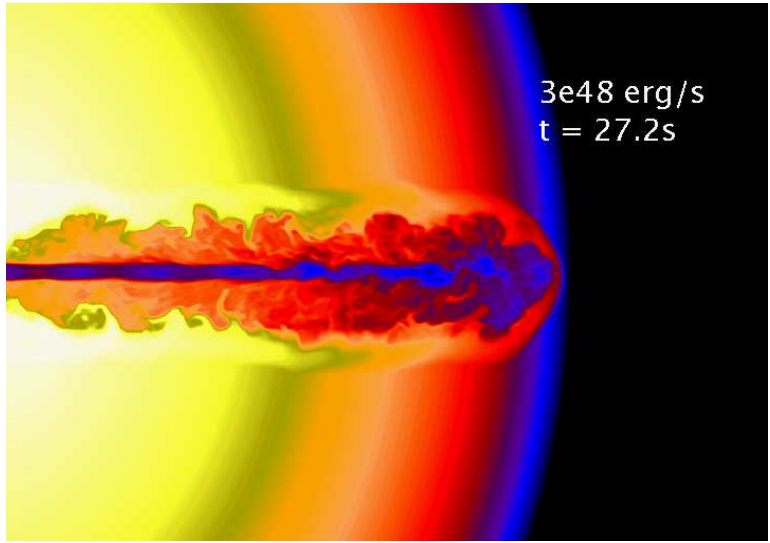}
\end{center}
 \caption{Three-dimensional calculation by Weiqun Zhang of a
   relativistic jet of $3 \times 10^{48}$ erg s$^{-1}$ introduced at
   $1 \times 10^{10}$\,cm in a 15\,\Msun \ Wolf-Rayet presupernova
   star of radius $8 \times 10^{10}$\,cm. Plotted is the logarithm of
   the density as the jet nears the surface. The jet took much longer
   to reach the surface than a similar jet with power $3 \times
   10^{50}$ erg s$^{-1}$ studied by Zhang
   et~al.\ (2004)\protect\nocite{Zha04} and was less stable.  After
   break out, the jet eventually becomes more stable as an opening is
   cleared by the relativistic flow. }
\end{figure}

\subsection{Weak jets and suffocated jets} 

There is a minimum power that the central engine must provide for a
relativistic jet to escape in a reasonable time, even in a
hydrogen-stripped massive star.  Figure~1 (Woosley \& Zhang
2007)\nocite{Woo07b} shows the density structure just as a jet of $3
\times 10^{48}$ erg s$^{-1}$ jet erupts from the surface of the star
27 s after initiation. Two other higher energy jets, 0.3 and $3 \times
10^{50}$ erg s$^{-1}$ took 15 s and 7 s respectively.  While a
relativistic jet of arbitrarily low power will eventually break out of
any star, the stellar structure will have changed and the central
engine may have turned off. Highly relativistic jets might, in some
cases, {\it never} emerge from the star (``failed GRB''). Such jets
could, however, deliver considerable energy focused on a small solid
angle of the stellar surface. The same fate would await a powerful jet
that did not maintain its orientation at the center to within a few
degrees (Zhang et~al.\ 2004)\nocite{Zha04}. Depending on how close to
the surface the jet makes it before dissipating, one could have either
a mildly asymmetric supernova with essentially no relativistic ejecta,
or a weak LSB powered by shock breakout (Tan
et~al.\ 2001)\nocite{Tan01} or collision with circumstellar matter
(Wang et~al.\ 2007; Katz et~al.\ 2009)\nocite{Wan07,Kat09}. Perhaps
this occurred in GRB\,980425.

\subsection{Shock breakout} 

Even the breakout of a spherically symmetric shock in an ordinary
(10$^{51}$ erg) Type Ib supernova will produce a soft X-ray transient
(Colgate 1968; Blinnikov et~al.\ 2003)\nocite{Col68,Bli03}. Such a
transient was observed (Soderberg et~al.\ 2008)\nocite{Sod08} for Type
Ib SN 2008D. Its duration, $\sim400$ s in the 0.3 - 10 keV band, was
much longer than anticipated on the basis of previous
models. Soderberg et~al.\ (2008) attributed the long time scale to
circumstellar interaction. Others invoked a mildly relativistic jet
energized by black hole formation (Mazzali
et~al.\ 2008)\nocite{Maz08}, wind interaction, or an extended radius
for the progenitor (Chevalier \& Fransson
2008)\nocite{Chev08}. Evidence for the latter interpretation comes
from recent studies (Yoon et~al.\ 2010)\nocite{Yoo10} that show Type
Ib progenitors in binaries have considerably larger radii than
previously expected. Hopefully, a large number of these events will be
discovered in the supernova surveys planned for the next decade. While
not GRBs themselves, they may help to elucidate the physics of shock
breakout in compact progenitors similar to those where GRBs occur.

\subsection{``Gamma-ray bursts'' with low luminosity and long
duration} 

If only the matter near the surface of the star has enough angular
momentum to form a disk outside a black hole, a longer fainter
transient could still be produced. The physics of jet production would
be similar to other collapsars, but the collapse time scale would be
longer. Most of the binary models proposed for LSBs give a surface
with high angular momentum, but do not speed up the core
(\Sect{rotation}). Chemically homogeneous evolution including magnetic
torques frequently gives helium cores in which only the outer layers
can form a disk (Woosley \& Heger 2006)\nocite{Woo06b}. Even some
models that make red and blue supergiants have enough rotation in
their outer envelope to make a disk, unless the mass loss rate is high
(\Sect{pairsn}).  Despite a similar central engine, the emerging jets
might have different properties because of the lower accretion rate,
larger black hole mass, and the fact that the jet has little or no
star left to penetrate.  The most luminous of these transients, in
which roughly a solar mass accreted from the surface of a Wolf-Rayet
progenitor with about a solar radius, would have a duration in its
rest frame of several minutes and a luminosity similar to (weaker)
LSBs (e.g., Janiuk \& Proga 2008\nocite{Jan08}). In progenitors that
are giants however, the time scale could be days and the event might
resemble a blazar or ultra-luminous X-ray source more than an LSB (see
also \Sect{pairsn}). The total energy in the event would be comparable
to LSBs, $\sim10^{51}$ erg. Any accompanying supernova would be faint
(Lopez-Camara et~al.\ 2008)\nocite{Lop08} unless the star were a
giant.

\subsection{Off-axis phenomena} 

LSBs require relativistic jets for their production and this means
that what an observer sees will be strongly biased by their
location. For a jet with sharp boundaries, the emission intensity
falls off rapidly with viewing angle (Granot et~al.\ 2002, 2005;
Ramirez-Ruiz et~al.\ 2005)\nocite{Gra02,Ram05,Gra05}, but such sharp
boundaries are unrealistic. There will thus be emission as the cocoon
of the jet breaks through the stellar photosphere (e.g., Ramirez-Ruiz
et~al.\ 2002; Zhang et~al.\ 2003, 2004\nocite{Ram02,Zha03,Zha04}) and
due to the interaction of the cocoon with the circumstellar
medium. Such interactions might give rise to X-ray flashes (Chapter 4;
Pe'er et~al.\ 2006)\nocite{Pee06} or enhanced X-ray emission in an LSB
(Peng et~al.\ 2005; Butler 2007; Ghisellini
et~al.\ 2007)\nocite{Pen05,But07,Ghi07}.

\subsection{Pair-instability supernovae} 
\lSect{pairsn}

SN 2006gy and SN 2007bi have demonstrated that stars with helium cores
of at least 45\,\Msun, and perhaps over 100\,\Msun \ are dying, even
in the modern Universe (Smith et~al.\ 2007; Woosley et~al.\ 2007;
Gal-Yam et~al.\ 2009)\nocite{Smi07,Woo07,Gal09}. Such stars encounter
the pair-instability, first as a violent pulsational instability for
helium cores in the mass range M$_{\rm He}$ = 40 - 60\,\Msun \ (main
sequence masses from 95 to 130\,\Msun), and then as a violent
explosive instability that disrupts the entire star for M$_{\rm He}$ =
60 - 133 \Msun \ (main sequence masses from 130 to 260\,\Msun). Above
260\,\Msun \ black holes are formed after helium depletion. These
numbers are for non-rotating stars of constant mass (Woosley
et~al.\ 2007; Heger \& Woosley 2002; Heger
et~al.\ 2003)\nocite{Woo07,Heg02,Heg03}. For rotating stars, the
relevant mass ranges are reduced.

Both pulsational pair instability supernovae (\Sect{ppsn}) and stars
over 260\,\Msun \ are likely to make black holes. Fryer et
al.\ (2001)\nocite{Fry01} studied the collapse of a 300\,\Msun \ star,
but assumed a rotating progenitor in which magnetic torques had been
ignored in the pre-collapse evolution. Even with an extended hydrogen
envelope, the helium core in this case maintained a specific angular
momentum, $j \sim 10^{18}$ cm$^2$ s$^{-1}$, enough to form an
accretion disk before the entire helium core went inside the event
horizon. Based upon an estimate of the viscous time scale for the
disk, Fryer et al.\ (2001)\nocite{Fry01} estimated a transient of
order 10\,s duration in the rest frame, but speculated that it could
be much longer.

No calculations have been published of progenitor evolution in this
mass range that include magnetic torques, but unpublished studies of a
250\,\Msun \ model with metallicity 10$^{-4}$ solar (Woosley
2010)\nocite{Woo09b} show that a variety of outcomes are possible. The
results depend upon the rotation rate assumed and the way magnetic
torques and mixing at the boundaries of convective regions are
treated. For the same physics as in Woosley \& Heger
(2006)\nocite{Woo06b}, three models were studied with total angular
momentum $J = 0.75, 1.0$ and $1.5 \times 10^{54}$\,erg\,s
corresponding to equatorial rotation speeds on the main sequence of
170\,km\,s$^{-1}$, 220\,km\,s$^{-1}$, and 305\,km\,s$^{-1}$. These
stars died with helium core masses - 142, 166, and 222\,\Msun, so all
three made black holes.  However, all stars were supergiants in their
late stages and, because of the magnetic torques, their helium cores
ended up rotating about an order of magnitude slower ($j \sim 10^{16 -
  17}$ cm$^2$ s$^{-1}$) than in the previous study. Consequently the
entire helium core and most of the hydrogen envelope would collapse
into the hole promptly in all three cases without making a
disk. However, the angular momentum extracted from the core increased
that of the hydrogenic envelope. The specific angular momentum in the
outer 20 -- 40\,\Msun \ exceeded 10$^{18}$ cm$^2$ s$^{-1}$ in all
models. In the outer 10\,\Msun \ it even exceeded 10$^{19}$ cm$^2$
s$^{-1}$, sufficiently high to form a disk.

This matter is at large radius and low density, so its free fall time
scale is roughly 10$^4$ to 10$^5$ s. If it is not lost from the star
prior to the collapse, the outer solar mass, with $j \sim 10^{20}$
cm$^2$ s$^{-1}$ would become supported by rotation at $3 \times
10^{11}$ cm with an orbital period of $\sim$1000 s. Depending upon
disk viscosity, it might take a day to fall in. An accretion rate of 1
\Msun \ day$^{-1}$ with an efficiency for mass to energy conversion of
10\% would give a luminosity of ~10$^{48}$ erg s$^{-1}$, even without
beaming.  A similar model was recently discussed by Komissarov and
Barkov (2010)\nocite{Kom10}.

\subsection{Pulsational pair supernovae} 
\lSect{ppsn}

Pulsational pair-instability supernovae eject solar masses of material
in repeated supernova-like outbursts with intervals that can span days
to centuries (Woosley et~al.\ 2007)\nocite{Woo07} before finally
collapsing to a neutron star or a black hole.  The variable intervals
and masses ejected allow the possibility of a wide range of observed
phenomena. Because of their short lifetimes, the iron core in these
stars, which may exceed 2 or even 3\,\Msun, remains rapidly
rotating. Black hole formation seems likely, but even if it is
avoided, the rapid rotation would produce a millisecond magnetar.  In
most cases, any jet produced would dissipate its relativistic energy
while still buried in optically thick ejecta. In some cases, though,
primarily those on the lighter end of the mass scale, the interval
between shell ejection is long enough that an LSB could happen inside
a previous supernova. The ``supranova'' model (Vietri \& Stella
1998)\nocite{Vie98} in which an LSB can happen inside of a supernova
that occurred years before is, in some sense, still with us.

\subsection{Short hard bursts and bursts at high redshift}
\lSect{hiz}

One of the major goals of GRB astronomy has been the discovery of
events with very high redshifts. This is motivated both by the desire
to find evidence for massive star death in the early Universe, and by
the belief that the first and second generation stars in the Universe
may have had unusual properties. In particular, they might have had
high mass, low metallicity and rapid rotation, just the sorts of
things that favor LSBs. Earlier this decade, it was hoped that one
might find luminous GRB counterparts to pair-instability supernovae
(Fryer et~al.\ 2001)\nocite{Fry01}. That still could happen, but as
argued in \Sect{pairsn}, that sort of star death now seems more likely
to produce long transients with low luminosity.

What has been found instead is confusing. GRB\,050904, one of the more
distant bursts discovered (z= 6.3) had a duration ($T_{90}$) of 220 s
(Cusumano et~al.\ 2006, 2007)\nocite{Cus06,Cus07} with flaring
activity that lasted one to two hours. The total energy in
relativistic ejecta was $\sim$10$^{52}$ erg (Frail
et~al.\ 2006)\nocite{Fra06}.  In addition, GRB\,050730 at $z = 3.969$,
GRB\,050505 at $z = 4.27$, and GRB\,050814 at $z = 5.3$ were all
exceptionally long lasting and among the brightest GRBs ever observed
(Cusumano et~al.\ 2007)\nocite{Cus07}.  So the case was starting to
look good for very energetic LSBs at high redshift (\Sect{Zmod}).  But
then came GRB\,080913 (Greiner et~al.\ 2009)\nocite{Gre09} and
GRB\,090423 (Tanvir et~al.\ 2009; Salvaterra
et~al.\ 2009)\nocite{Tan09,Sal09}, the two most distant GRBs to date,
at redshifts 6.7 and 8.2 respectively. By their observed durations,
these events would be classified as LSBs, but it was noted that in
their frame, the duration ($T_{90}$) of each was less than 2\,s and
their prompt emission was hard; therefore it was argued that they
belong to the SHB class.  However, one should remember that the
LSB/SHB classification scheme is based on observer-frame duration
distribution, according to which they are clearly LSBs.  These GRBs
also exhibited post-burst X-ray flaring activity with energy
comparable to the initial burst. Indeed, a more general discussion is
developing about a group of bursts, near and far, that straddle the
boundary between LSBs and SHBs according to various old, and new,
criteria (Perley et~al.\ 2009; Zhang
et~al.\ 2009)\nocite{Per09,Zha09}.

One possibility is that we are witnessing the break down, or at least
substantial modification of the highly successful internal-shock model
(Rees \& M\'esz\'arose 1994)\nocite{Ree94} for GRBs. It has become
common, without really good cause even in the internal-shock model, to
associate the duration of a burst, especially $T_{90}$, with the
activity cycle of its central engine, i.e., merging compact objects
for short bursts and massive star death for long ones, with a dividing
line at 2\,s.  Perhaps in the very high redshift bursts and other
energetic short bursts, internal shocks are less efficient or have a
different characteristic time scale (Zhang
et~al.\ 2009)\nocite{Zha09}. Zhang et~al.\ (2003)\nocite{Zha03} have
also discussed the possibility of producing short bursts in massive
stars by an {\it external} shock mechanism in the medium just outside
the star, and Waxman et~al.\ (2003)\nocite{Wax03} have also discussed
SHBs as a break out phenomenon. Finally, the emission of a Poynting
flux jet as it erupts from the surface of a star has yet to be
calculated, and may be quite different from the later emission after
the jet has cleared a broad passage through the star.  If the prompt
emission is really just some sort of break out transient, the distant
bursts from long ago may end up telling us about interesting
variations in the massive stellar progenitor properties - how they
collimate and modulate jets and their circumstellar densities - but
may not help to discriminate the ``central engine''. An exception
would be if an event were discovered that had a total energy beyond
what a magnetar can provide.

\section{Possible model diagnostics}

\subsection{The maximum energy of the explosion}
\lSect{maxe}

The minimum rotational period of a rigidly rotating neutron star is
near 1\,ms (Lattimer \& Prakash 2007)\nocite{Lat07}. Because of the
magneto-rotational instability, a differentially rotating neutron star
is not likely to remain so for long, but this final rapid rotation
rate can only be achieved after a several second long Kelvin-Helmholtz
evolution of the neutron star.  During this time, magnetic torques and
jets may appreciably dissipate the rotation (Dessart
et~al.\ 2008)\nocite{Des08}. As a result, the maximum rotational
energy available to explode a supernova and power an LSB in the
magnetar model is probably only a few times $10^{52}$ erg (Ott
et~al.\ 2006; Burrows et~al.\ 2007b)\nocite{Ott06,Bur07b}, and Adam
Burrows, private communication).  This is smaller than the kinetic
energies inferred in some models for ``hypernovae'' (e.g., Maeda \&
Nomoto 2003\nocite{Mae03}), but these limits from one-dimensional
models are not precise. Collapsars can, in principle, provide a much
larger energy, up to a substantial fraction of the rotational energy
of a Kerr hole or the accretion energy of several solar masses, both
$\sim10^{54}$ erg. In the common case, a smaller value is expected.
The highest current lower limits are thus close to the maximum allowed
in the magnetar model (\Sect{energy}), but are not yet in conflict.

\subsection{$^{56}$Ni and supernovae}
\lSect{ni56}

The magnetar model and collapsar model make their $^{56}$Ni in
different ways - the magnetar by a shock wave, the collapsar by a disk
wind. Both models are still rudimentary regarding the amount of
$^{56}$Ni that is made. In the magnetar model, the same engine must
produce: i) enough energy in a directed relativistic form, over an
extended period of time, to power the burst; ii) enough nearly
isotropic energy to blow up the star; and iii) at least occasionally,
a few tenths of a solar mass of $^{56}$Ni to make the supernova
bright. The latter requires depositing at least several times
10$^{51}$ erg in a large solid angle in much less than one second in
order to heat sufficient matter above $5 \times 10^9$ K.  So far, the
published magnetar models relegate the $^{56}$Ni problem to some
precursor explosion that sets up the circumstances for the magnetar to
operate.

Still, it might be possible to thread this needle. There are several
relevant time scales in the magnetar model. First is the time required
to dissipate the {\it differential} rotation of the proto-neutron
star. This involves magnetic field generation and instabilities and
could be short compared with the Kelvin-Helmholtz timescale (Dessart
et~al.\ 2008)\nocite{Des08}. Depending upon how and where this energy
is dissipated, starting from a rigidly rotating iron core, there is
enough energy in differential rotation alone to power the supernova
and make the necessary $^{56}$Ni. The other time scales are the
Kelvin-Helmholtz time scale, during which the neutron star reaches its
terminal rotation speed, and the uncertain time scales for spin down
and accretion. The possibility of accretion is often overlooked in the
magnetar model. If the proto-neutron star has too much angular
momentum to be a millisecond pulsar, it may shed the excess in a disk
that it later accretes on a viscous time scale. At that point, the
model would resemble both the outcome of merging neutron stars and the
collapsar model, except that the disk mass would be small compared
with the latter.

Given the possibilities, a magnetar model can probably be evolved that
provides the necessary $^{56}$Ni yield. Alternatively, the brightness
of the supernova may not be due to radioactivity (Kasen \& Bildsten
2010; Woosley 2009)\nocite{Kas09,Woo09}. Either way, additional data
on the SN-LSB connection and more realistic theoretical models will be
helpful. Given the difficulty making $^{56}$Ni in the first place, a
LSB without a bright supernova should not be difficult to arrange in
the magnetar model, but an LSB with {\it no} supernova would be
puzzling.

The collapsar model can also, in principle, make large amounts of
$^{56}$Ni and explode the star violently while still having plenty of
enduring power for the burst. However, this all relies on the
hydrodynamics of the disk wind which has been poorly explored in
realistic models. It is certainly possible for it to make very
little. The $^{56}$Ni that is made could accrete onto the black hole
during a ``fall-back'' phase. $^{56}$Ni production might also be low in that
variety of collapsar where the black hole itself is a result of
fall-back (Fryer et~al.\ 2007)\nocite{Fry07}.

\subsection{Continued activity of the central engine}

{\it Swift\/} has seen ``flares'' of hard X-ray emission occurring
hundreds to thousands of seconds after the main burst (Falcone
t~al.\ 2007; Chincarini et~al.\ 2007)\nocite{Fal07,Chi07}. These
resurgences of emission, which can sometimes contain as much energy as
the principal burst (but more typically $\sim$10\%), are generally
taken to be evidence of continuing activity of the central
engine. Flare durations are often quite short compared with the
elapsed time since the burst onset and their spectrum is harder than
that of the underlying continuous emission.

Dai et~al.\ (2006)\nocite{Dai06} have suggested that late flares can
be explained by magnetic reconnection events driven by the breakout of
magnetic fields from the surface of differentially rotating
millisecond pulsars (see also Klu\'zniak \& Ruderman
1998\nocite{Klu98}). Their model was suggested in the context of short
hard bursts, but might apply to the magnetar model for long bursts as
well. Giannios (2006)\nocite{Gia06} also suggested reconnection, but
in the ejecta themselves, placing the origin far from the
source. However, both these explanations must confront the fact that
the flares have observational properties that are very similar to the
prompt emission (Krimm et~al.\ 2007; Margutti
et~al.\ 2010)\nocite{Kri07,Mar10}.

An alternate explanation, and one perhaps more favorable for the more
energetic flares, is transient accretion (King et~al.\ 2005; Lazzati
et~al.\ 2008; Kumar et~al.\ 2008)\nocite{Kin05,Laz08,Kum08}.  Fallback
can continue from the supernova that accompanies the LSB at an
appreciable rate for hours after the initial burst (MacFadyen
et~al.\ 2001; Kumar et~al.\ 2008; Zhang
et~al.\ 2008)\nocite{Mac01,Kum08,Zha08} and may be
unsteady. Unpublished 2D collapsar calculations by Weiqun Zhang,
sometimes show an oscillatory cycle in which a weak equatorial
explosion is launched by energy released in the disk. Material moves
out for a time, shutting off the accretion, then falls back in again
re-establishing the disk. The natural time scale for these cycles is
roughly the dynamic time for the helium core, $\sim100$ s, but wide
variations are possible. One might expect such oscillations to be more
prominent in very massive stellar cores that are hard to unbind.

\noindent {\bf Acknowledgements}

\noindent This work has been supported by the NASA Theory Program
(NNG05GG08G and NNX09AK36G). The author thanks Chryssa Kouveliotou and
Ralph Wijers for their perseverance in assembling this volume and
Ralph for a careful reading and editing of the manuscript.  Adam
Burrows and Weiqun Zhang provided important unpublished
details of their models.

\newcommand{\apj}{{\it ApJ,}}
\newcommand{\aj}{{\it AJ,}}
\newcommand{\apjl}{{\it ApJ,}}
\newcommand{\apjs}{{\it ApJS,}}
\newcommand{\aap}{{\it A\&A,}}
\newcommand{\araa}{{\it ARA\&A,}}
\newcommand{\mnras}{{\it MNRAS,}}
\newcommand{\nat}{{\it Nat,}}
\newcommand{\physrep}{{\it Phys.\ Rept.,}}
\newcommand{\apss}{{\it Ap.\ \& Spac.\ Sci.,}}
\newcommand{\prd}{{\it PRD,}}
\newcommand{\rmp}{{\it Rev.\ Mod.\ Phys.,}}

\end{document}